\documentclass[twocolumn,showpacs,preprintnumbers,prb,fleqn]{revtex4}
\usepackage{graphicx}
\usepackage{dcolumn}
\usepackage{bm}

\begin{document}

\title{Magnetophonon resonance in high density, high mobility quantum well systems}
\author{C. Faugeras, D. K. Maude and G. Martinez}
\affiliation{Grenoble High Magnetic Field Laboratory, Max Planck Institut f\"{u}r Festk%
\"{o}rperforschung\\
and Centre National de la Recherche Scientifique, BP 166, 38042
Grenoble Cedex 9, France.}
\author{L. B. Rigal and C. Proust}
\affiliation{Laboratoire National des Champs Magn\'{e}tiques
Pulses, \\143, avenue de Rangueil, F31432, Toulouse, Cedex 4,
France}
\author{K. J. Friedland, R. Hey and K. H. Ploog}
\affiliation{Paul Drude Institut f\"{u}r Festk%
\"{o}rperelektronik, Hausvogteiplatz 5-7, D-10117 Berlin, Germany}
\date{\today }

\begin{abstract}
We have investigated the magnetophonon resonance (MPR) effect in a
series of single GaAs quantum well samples which are symmetrically
modulation doped in the adjacent short period AlAs/GaAs
superlattices. Two distinct MPR series are observed originating
from the $\Gamma$ and X electrons interacting with the GaAs and
AlAs longitudinal optic (LO) phonons respectively. This confirms
unequivocally the presence of X electrons in the AlAs quantum well
of the superlattice previously invoked to explain the high
electron mobility in these structures (Friedland et al. Phys. Rev.
Lett \textbf{77},4616 (1996))
\end{abstract}

\maketitle

Ultra high conductivity ($\sigma=n e \mu$) two-dimensional
electron gas (2DEG) systems have important applications in low
noise, high frequency devices. Increasing the doping level, in
order to increase the carrier density (n), often reduces $\sigma$,
due to the significant decrease in mobility ($\mu$) which is a
direct result of the increased scattering from the remote ionized
donors. For modulation doped 2-dimensional (2D) structures based
on the GaAs/AlAs system, the use of short period superlattices for
the incorporation of the remote doping layer has been shown to
lead to significant enhancement of the mobility for high density
($\sim 1 \times 10^{12}$ $cm^{-2}$) 2DEG samples\cite{Friedland}.
The achieved mobilities, up to $\sim 3 \times 10^{6}$
$cm^{2}V^{-1}s^{-1}$, have been attributed to a screening of the
Coulomb potential of the remote dopants by localized X-electrons
in the AlAs quantum well adjacent to each Si $\delta$-doped GaAs
layer in the short period superlattice. This hypothesis was based
on self consistent calculation of the space charge distribution in
the structure and capacitance-voltage measurements which indicated
the presence of carriers outside the center GaAs quantum well and
near to the Si $\delta$-doping region.

The expected Bohr radius of the X-electrons ($a_B\approx 2-3$
$nm$), together with the nominal distance of $1.7$ $nm$ from the
$\delta$ doped layer means that the X-electrons can efficiently
screen the ionized Si-donors atoms whose average separation is
$\approx 8-9$ $nm$. The X-electrons are effectively localized at
low temperatures ($T\leq 10-20$ $K$) and do not contribute to
transport (absence of parallel conduction). However, at higher
temperature, the X-electrons should become delocalized. An
unmistakable signature of a contribution of the X-electrons to
transport would be the observation of a magnetophonon resonance
(MPR) in the electrical transport properties. This resonant
scattering with longitudinal optic (LO) phonons in the presence of
an applied magnetic field is well known in both
bulk\cite{Puri,Firsov} and 2D semiconductors\cite{Tsui,Kido,Mori}.

The main result presented in this work is the observation, of two
distinct MPR series in our samples, which due to the very
different effective masses and phonon energies, we can identify
with $\Gamma$ and X-electrons in the GaAs and AlAs quantum wells
respectively. The observation of the AlAs MPR clearly validates
the important role played by X-electrons in screening the long
range potential of the remote ionized Si-donors.

The four different structures investigated here were grown by
solid source molecular beam epitaxy on semi-insulating (001) GaAs
substrates. The barriers consist of two short period superlattices
each with 60 periods of 4 monolayers of AlAs and 8 monolayers of
GaAs. Carriers are introduced into the central $10-13$ $nm$ GaAs
quantum well by a single Si $\delta$-doping sheet with a
concentration of $2.5\times10^{12}$ $cm^{-2}$ placed in a GaAs
layer of each short period superlattice. Full details of the
growth protocol are available elsewhere\cite{Friedland}. The
sample properties are summarized in Table \ref{tab:Table1}. Hall
bars were fabricated using conventional photolithography and
chemical wet etching. Ohmic contacts were formed by evaporating
AuGeNi followed by an anneal at $450^{o}C$.

For the magneto-transport measurements the samples were placed on
a rotation stage in a variable temperature cryostat installed on a
water cooled, 28T, 20MW dc magnet. Transport measurements were
performed using conventional phase sensitive detection at $10.66$
$Hz$ with a current of $1-5$ $\mu A$. For the very high field
measurements a 60 $T$ pulsed magnet was used with conventional dc
detection ($I=100$ $\mu A$) and a low pass preamplifier to limit
the transient component of the signal induced by the magnetic
field pulse. In both cases the longitudinal sample resistance
$R_{xx}$ and Hall resistance $R_{xy}$ were simultaneously measured
as a function of magnetic field (B). The tilt angle of the sample
with respect to the applied magnetic field can be precisely
determined from $R_{xy}$ which depends only on the perpendicular
magnetic field ($B_{\perp}$).

\begin{table}
\caption{\label{tab:Table1}Electron density n$_s$ and mobility
$\mu$ measured at $T=300$ $mK$ for different samples with quantum
wells of width L$_{QW}$ at a distance d$_{\delta}$ from the remote
$\delta$-doping layers.}
\begin{ruledtabular}
\begin{tabular}{ccccc}
 &$n_s$ &$\mu$&L$_{QW}$ & d$_{\delta}$\\
 \multicolumn{1}{c}Sample & (cm$^{-2}$)& (cm$^2$/V.s) & (nm) & (nm) \\
\hline
$1038$& $1.28\times10^{12}$& $1.14\times10^{6}$ & $10$ & $14$\\
$1201$& $9.4\times10^{11}$& $2.80\times10^{6}$ & $13$ & $22$\\
$1200$& $7.4\times10^{11}$& $2.20\times10^{6}$ & $13$ & $30$\\
$1416$& $6.4\times10^{11}$& $2.18\times10^{6}$ & $13$ & $34$ \\
\end{tabular}
\end{ruledtabular}
\end{table}

The application of a magnetic field quantizes the electron motion
and to a first approximation resonant absorption of an LO phonon
can occur whenever,
\begin{eqnarray}
\label{eq1} \hbar \omega_{LO}=N \hbar \omega_{c} = N \hbar e
B_{\perp}/m^{*}\nonumber
\end{eqnarray}
where $N=1,2,3...$, $\hbar \omega_{LO}$ is the LO phonon energy
and $m^{*}$ is the electron effective mass. The changed scattering
rate at resonance gives rise to oscillations in the
resistance\cite{Tsui,Kido} which are periodic in $1/B_{\perp}$.
The optimum temperature for observing MPR oscillations is a
compromise between having the lowest temperature possible in order
to reduce Landau level broadening and having the highest
temperature possible in order to have a large thermal phonon
population. Even under optimum conditions MPR oscillations are
weak\cite{Brummel87} with a typical amplitude $\delta R/R \sim 1
\%$.

\begin{figure}
\includegraphics[width=1.0\linewidth,angle=0,clip]{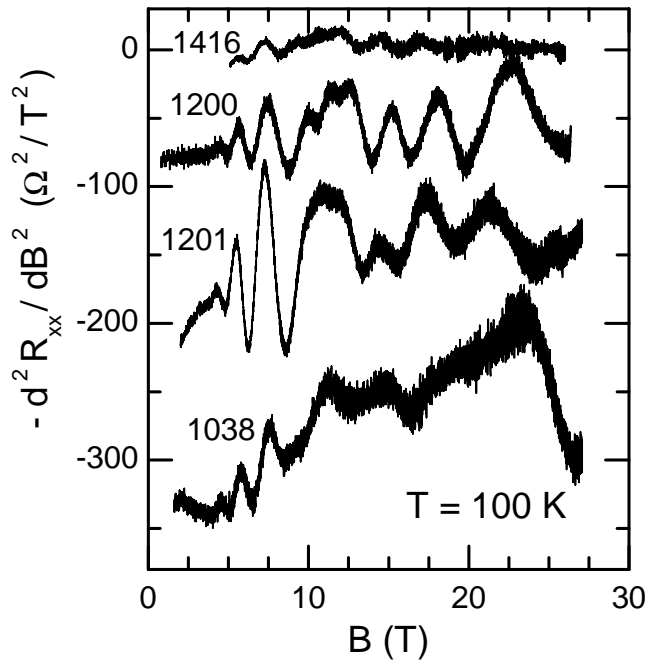}
\caption{\label{fig1} Second derivative of the longitudinal
resistance versus magnetic field measured at $\theta = 0$ and
$T=100K$ for the four samples investigated. Curves have been
shifted vertically for clarity}
\end{figure}

In order to extract the weak MPR oscillation from the
monotonically increasing background magnetoresistance we have
numerically calculated the second derivative $d^2R_{xx}/dB^2$.
Results for the four samples measured at $T=100$ $K$ (the optimum
temperature for our samples) and with the magnetic field applied
perpendicular to the layers ($\theta = 0$) are shown in
Fig.\ref{fig1}. All samples show more or less pronounced MPR
oscillations with two distinct series. A low magnetic field series
corresponding to the scattering of the $\Gamma$-electrons in the
center quantum well with GaAs LO phonons and a second series which
develops at higher magnetic fields ($B\gtrsim 10$ $T$)which we
identify with the scattering of X-electrons with AlAs LO phonons
in the AlAs layer next to each Si $\delta$-doping in the short
period superlattice. The fundamental magnetic field (corresponding
to $N=1$) for each series of oscillations can be obtained from
either an analysis of the period of the oscillations in $1/B$ or
from a Fourier transform of $d^2R_{xx}/dB^2$ versus $1/B$. Both
methods give identical results within experimental error. The
fundamental fields deduced for each sample (where possible) are
indicated in Table \ref{tab:Table2}.

\begin{table}
\caption{\label{tab:Table2}Fundamental magnetic fields for the MPR
oscillations in Fig.\ref{fig1}. The phonon energies have been
calculated assuming $m^*/m_e=0.072$ for the $\Gamma$-electrons in
GaAs and $m^*/m_e=0.21$ for the light (in plane) mass  X-electrons
in AlAs }
\begin{ruledtabular}
\begin{tabular}{ccccc}
 &$B_F(GaAs)$ &$B_F(AlAs)$&$\hbar\omega_{LO}(GaAs)$ & $\hbar\omega_{LO}(AlAs)$\\
 \multicolumn{1}{c}Sample & (T)& (T) & (meV) & (meV) \\
\hline
$1038$& $22.5\pm 0.7$& - & $36.7\pm 1.1$ & -\\
$1201$& $21.1\pm 0.6$& $89.2\pm 1.5$ & $34.4\pm 2.1$ & $49.9\pm 2.3$\\
$1200$& $22.9\pm 0.5$& $88.4\pm 2.3$ & $37.3\pm 2.0$ & $49.4\pm 2.8$\\
$1416$& $23.3\pm 1.2$& $89.5\pm 5.5$ & $38.0\pm 3.0$ & $50.1\pm 5.0$\\
\end{tabular}
\end{ruledtabular}
\end{table}

The AlAs series has a considerably higher fundamental magnetic
field due to both the larger effective mass of the X-electrons and
the larger LO phonon energy of AlAs. In order to take into account
non parabolicity it is necessary to use slightly heavier effective
masses than the band edge values\cite{Adachi} for GaAs
($m^*/m_e=0.067$) and AlAs ($m^*/m_e=0.19$). The calculated phonon
energies which are summarized in Table \ref{tab:Table2} have been
calculated assuming a mean value of $m^*/m_e=0.072$ for the
$\Gamma$-electrons in GaAs and $m^*/m_e=0.21$ for the light mass
(in plane) X-electrons in AlAs. The phonon energies found are in
good agreement with the currently accepted values for bulk GaAs
($\hbar\omega_{LO}(GaAs)=36.25$ $meV$) and bulk AlAs
($\hbar\omega_{LO}(AlAs)=50.09$ $meV$)\cite{Adachi}.

In 2D the coupling to phonon modes is expected to be modified,
with the electrons coupling to the interface (slab)
modes\cite{Brummel87} or dressed modes due to dynamic
screening\cite{Afonin} both of which are expected to have a
slightly reduced energy compared to bulk. Far infrared absorption
measurements at zero magnetic field in these samples confirm that
the LO-phonon energies are indeed close to the bulk
values\cite{Faugeras}. A precise determination of the phonon
energy, for which the effective mass should be precisely known, is
not the objective of this work.

To our knowledge, MPR has never been observed in bulk AlAs. In 2D
systems MPR has been reported due to scattering between quantum
well $\Gamma$ electrons and evanescent AlAs  barrier or interface
LO phonons in GaInAs/AlInAs multi-quantum wells\cite{Nicholas} or
AlGaAs/GaAs superlattice structures\cite{Noguchi}. MPR between
AlAs X-electrons and GaAs LO phonons has been
observed\cite{Ferrus} in 1D GaAs/AlAs quantum wires, and in
GaAs/AlAs superlattice samples when the magnetic field is applied
along the growth direction.  We stress here that it is not
possible to identify the high field MPR series with the scattering
between $\Gamma$ electrons (in the quantum well) and evanescent
AlAs barrier or interface LO phonons since using the light
$\Gamma$ mass $m^*/m_e=0.072$ would imply a scattering with a
phonon of energy $\sim 150$ $meV$ which far too large to be
physically meaningful. Therefore, the observation of the high
field MPR series clearly demonstrates the presence of X-electrons
located in the AlAs barrier which interact with AlAs LO phonons.

\begin{figure}
\includegraphics[width=1.0\linewidth,angle=0,clip]{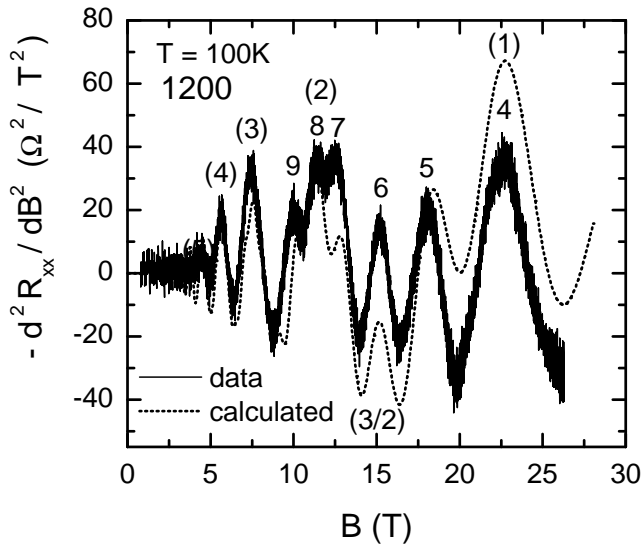}
\caption{\label{fig2} Comparison of the MPR oscillations
calculated using an exponentially damped cosine function and the
data for sample 1200. A linear background has been subtracted from
the raw data shown in Fig.\ref{fig1}. The minima are labelled to
indicate the resonance condition for the GaAs LO phonon series,
N=1, 2, 3, 4, 5 (in brackets) and for the AlAs LO phonon series,
N=4, 5, 6, 7, 8, 9.}
\end{figure}

In bulk GaAs MPR oscillations have been described by an
exponentially damped cosine series\cite{Stradling},
\begin{eqnarray}
\label{eq2} \Delta R_{xx}/R_{xx} \propto \cos (2\pi
\omega_{LO}/\omega_c)  \exp (-\gamma
\omega_{LO}/\omega_c)\nonumber
\end{eqnarray}
where $\gamma$ is a magnetic field independent damping factor. In
our case we can write
\begin{eqnarray}
\label{eq3} d^2R/dB^2\approx&-&A_1\cos(2\pi
\omega_{LO_1}/\omega_{c_1})
\exp (-\gamma_1 \omega_{LO_1}/\omega_{c_1})\nonumber\\
&-&A_2 \cos (2\pi\omega_{LO_2}/\omega_{c_2}) \exp
(-\gamma_2\omega_{LO_2}/\omega_{c_2})\nonumber
\end{eqnarray}
where the indices 1 and 2 refer to the relevant material
parameters for GaAs and AlAs respectively. The predicted behavior
is compared to the data for one of the samples in Fig.\ref{fig2}.
Here we have used the phonon energies for bulk GaAs
$\hbar\omega_{LO_1}=36.25$ $meV$ and AlAs
$\hbar\omega_{LO_2}=50.1$ $meV$ together with the relevant
effective masses $m^*/m_e=0.072$ for the $\Gamma$-electrons in
GaAs and $m^*/m_e=0.21$ for the light (in plane) mass X-electrons
in AlAs. The only adjustable fitting parameters are then $A_1=50$
$\Omega^{2}T^{-2}$, $A_2=100$ $\Omega^{2}T^{-2}$ and
$\gamma_1=\gamma_2=0.3$. The predicted behavior reproduces almost
exactly the observed period(s) and amplitude of the oscillations
except for a difference in amplitude at high magnetic field. This
can be attributed to a strong damping of the $N=1$ maximum and
$N=3/2$ minimum of the GaAs series. This series is strongly damped
at high magnetic fields due to self consistent corrections to the
density of states\cite{Leadley}. This is a result of the dominance
of elastic over inelastic (phonon) scattering at high magnetic
fields when the Landau levels are extremely narrow. The Landau
levels in the AlAs quantum well are significantly broader due to
the inevitably much lower mobility associated with the proximity
of Si $\delta$-doping layer. This suggests that it is the Landau
level broadening due to the presence of this short range
scattering which explains the robustness of the AlAs MPR series at
high magnetic field.

\begin{figure}
\includegraphics[width=1.0\linewidth,angle=0,clip]{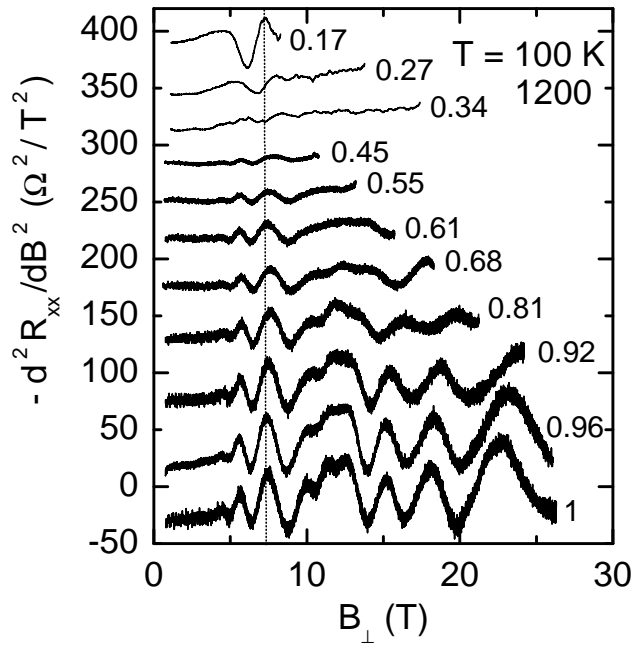}
\caption{\label{fig3} Second derivative of the longitudinal
resistance versus magnetic field measured at $T=100$ $K$ and for
different angles measured using static ($\cos (\theta) = 0.45-1$)
and pulsed ($\cos (\theta) = 0.17-0.34$) magnetic fields for
sample 1200. Curves have been shifted vertically for clarity. The
dotted vertical line at the magnetic field corresponding to $N=3$
for the GaAs series is drawn as guide to the eye.}
\end{figure}

In addition, rotation measurements have been performed to confirm
the 2D nature of both series. Typical results for sample 1200 are
shown in Fig.\ref{fig3} where $d^2R_{xx}/dB^2$ versus the
perpendicular component of magnetic field is plotted for a number
of different angles. With increasing tilt angle, for both series
the peak position do not shift with respect to $B_{\perp}$ as
expected for 2D systems for which the cyclotron energy depends
only on $B_{\perp}$. This is in agreement with the self consistent
calculations\cite{Friedland}, that the AlAs X-electrons are
localized in an AlAs quantum well and that there is no X-miniband
formation in the short period superlattice. The small shift of the
AlAs series to higher $B_{\perp}$ at large tilt angles is probably
due to the anisotropic mass of the X-valleys in AlAs.

For a given $B_{\perp}$, tilting the sample increases the in plane
component of the magnetic field ($B_{\parallel}$) which induces a
mixing between different electrical sub-bands. In GaAs/AlGaAs
heterojunctions this leads to a drastic reduction in the amplitude
of the MPR oscillations\cite{Brummel88}. The effect of
$B_{\parallel}$ on the MPR oscillations is less marked in quantum
well samples\cite{Nicholas} due to the much larger energy
separation of the electrical subbands. In Fig.\ref{fig3} the high
field AlAs series is suppressed for tilt angles above $47^o$
($\cos(\theta)=0.68$). The low field GaAs series survives at
almost all angles and even apparently regains in amplitude at very
high tilt angles ($\cos(\theta)\leq 0.17$). It is not evident that
this reemergence should be associated with MPR. At high tilt
angles it is almost certainly not a good approximation to treat
the system as being 2-dimensional since the magnetic length
$\ell_B = \sqrt{\hbar/eB}$ is already significantly less than the
width of the GaAs quantum well.

In conclusion, a series of high density 2DEG samples designed to
reduce remote impurity scattering have been investigated using
magnetophonon resonance. Two distinct MPR series are observed, one
originating from $\Gamma$ electrons interacting with GaAs
LO-phonons in the center GaAs quantum well, and one originating
from X-electrons interacting with AlAs LO phonons in an AlAs
quantum well adjacent to the Si $\delta$-doping in the short
period superlattice situated either side of the center GaAs
quantum well. This result clearly demonstrates the presence of
X-electrons in the AlAs quantum well which have
previously\cite{Friedland} been invoked to explain the enhanced
screening of the remote impurity potential required to explain the
very high mobilities achieved in these high carrier density
samples.

\begin{acknowledgments}
We thank A.Riedel and H.Kostial for help in sample preparation.
\end{acknowledgments}

\bigskip

\bibliography{FaugerasMPR}

\end{document}